# Residual orbit of BC bump for CSNS/RCS

**Ming-Yang Huang**[1,2] ·**Sheng Wang**[1,2] ·**Shouyan Xu**[1,2]

**Abstract**

**Background**  In a beam cycle, there is a chicane bump (BC bump) in the injection region of the China Spallation Neutron Source (CSNS). It is a core part of the injection system and the important guarantee that the Linac beam injecting into the rapid cycling synchrotron (RCS). During the beam commissioning, the residual orbit of BC bump, i.e. the closed-orbit distortion (COD) due to the imperfect BC bump, was an important problem which can affect the orbit correction and beam loss control of the RCS.

**Purpose** The purpose is to develop a suitable method to judge and correct the residual orbit of BC bump.

**Methods** According to the physics design, there is no residual orbit of BC bump in theory. However, if the residual orbit of BC bump exists in the actual operation, the circular beam orbit outside the injection region would be affected. By comparing the circular beam orbit outside the BC bump region in theory and that measured in the actual operation, whether the residual orbit of BC bump exists can be judged. By using the three auxiliary windings of BC magnets, the residual orbit of BC bump can be corrected.

**Results** The numerical simulation results showed that the measurement and correction methods had worked well and they can be applied to the beam commissioning of CSNS/RCS. The measurement results in the machine study showed that the residual orbit of BC bump is not large and can be reduced by the three auxiliary windings of BC magnets.

**Conclusion**  The methods to judge and correct the residual orbit of BC bump worked well and had been applied to the beam commissioning of CSNS/RCS.

## Introduction

The China Spallation Neutron Source (CSNS), whose technology acceptance had been received in March, 2018, is a multidisciplinary platform [1, 2]. Its accelerator consists of an 80 MeV H$^-$ Linac which is upgradable to 300 MeV and a 1.6 GeV rapid cycling synchrotron (RCS) with a repetition rate of 25 Hz which accumulates an 80 MeV injection beam, accelerates the beam to the designed energy of 1.6 GeV and extracts the high energy beam to the target. The design goal of beam power is 100 kW and capable of upgrading to 500 kW [3].

Ming-Yang Huang

huangmy@ihep.ac.cn

1 Institute of High Energy Physics, Chinese Academy of Sciences, Beijing 100049, China

2 Dongguan Institute of Neutron Science, Dongguan 523808, China

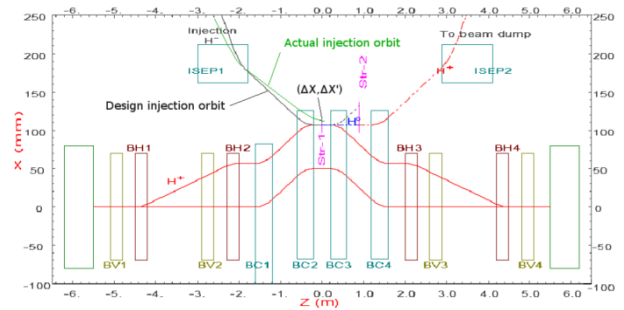

**Fig. 1** Layout of CSNS injection system

In order to reduce the beam loss caused by the space charge effects, the phase space painting in the position was used in both horizontal and vertical planes for CSNS [4, 5]. Fig. 1 shows the layout of CSNS injection system. It can be found that there are three kinds of orbit bumps: a horizontal chicane bump (four dipole magnets, BC1-BC4) in the middle for an additional closed-orbit shift of 60 mm; a horizontal bump (four dipole magnets, BH1-BH4) used for painting in the horizontal plane; a vertical bump

(four dipole magnets, BV1-BV4) used for painting in the vertical plane.

During the beam commissioning of CSNS [6-8], the residual orbit of BC bump, i.e. the closed-orbit distortion (COD) due to the imperfect BC bump, can affect the orbit correction and beam loss control of the RCS which need to be studied in detail. In the following sections, the methods to measure and correct the residual orbit of BC bump will be given and applied to the beam commissioning. The problem of the residual orbit of BC bump will be satisfactorily solved.

## Basic method

Because of the compact space and vacuum box aperture in the injection region, there is a horizontal chicane bump which gives an additional closed-orbit shift of 60 mm. According to the physics design, there is no residual orbit of BC bump in theory. However, if the residual orbit of BC bump exists in the actual operation, the circular beam orbit outside the injection region would be affected and the orbit correction of the RCS would be less precise. The residual orbit of BC bump needs to be corrected in time.

## Measurement method to the residual orbit of BC bump

In order to measure the residual orbit of BC bump, four beam position monitors (BPMs) near the injection region (R4BPM02, R4BPM01, R1BPM01, R1BPM02) should be used. After actual measurement, the COD data of these four BPMs can be given

In theory, by using the measurement data of R4BPM01 and R4BPM02 and the transfer matrix without BC chicane bump, the beam position and slope at R1BPM01 can be calculated. Defining $(x_1, x'_1)$ and $(x_2, x'_2)$ as the transverse phase space coordinates of the circular beam at R4BPM02 and R4BPM01, $x_1$ and $x_2$ can be measured in the machine study. The transfer relationship from R4BPM01 to R4BPM02 can be given as

$$\begin{pmatrix} x_1 \\ x'_1 \end{pmatrix} = \begin{bmatrix} \sqrt{\frac{\beta_1}{\beta_2}} \times (\cos \Delta\Psi_{12} + \alpha_2 \times \sin \Delta\Psi_{12}) & \sqrt{\beta_1 \times \beta_2} \times \sin \Delta\Psi_{12} \\ \frac{(\alpha_2 - \alpha_1) \times \cos \Delta\Psi_{12} - (1 + \alpha_1 \times \alpha_2) \times \sin \Delta\Psi_{12}}{\sqrt{\beta_1 \times \beta_2}} & \sqrt{\frac{\beta_2}{\beta_1}} \times (\cos \Delta\Psi_{12} - \alpha_1 \times \sin \Delta\Psi_{12}) \end{bmatrix} \begin{pmatrix} x_2 \\ x'_2 \end{pmatrix}, \quad (1)$$

where $(\alpha_1, \beta_1)$ and $(\alpha_2, \beta_2)$ are the twiss parameters at R4BPM02 and R4BPM01, $\Delta\Psi_{12} = \psi_1 - \psi_2$ while $\psi_1$ and $\psi_2$ are the phases at R4BPM02 and R4BPM01. By using Eq. (1), the beam slope at R4BPM01 can be calculated:

$$x'_2 = \frac{x_1 - \sqrt{\frac{\beta_1}{\beta_2}} \times (\cos \Delta\Psi_{12} + \alpha_2 \times \sin \Delta\Psi_{12}) \times x_2}{\sqrt{\beta_1 \times \beta_2} \times \sin \Delta\Psi_{12}}. \quad (2)$$

By using the transfer matrix without BC chicane bump, defining $(x_{3t}, x'_{3t})$ as the transverse phase space coordinates of the circular beam at R1BPM01 in theory, the transfer relationship from R4BPM01 to R1BPM01 can be given as

$$\begin{pmatrix} x_{3t} \\ x'_{3t} \end{pmatrix} = \begin{bmatrix} 1 & L_{23} \\ 0 & 1 \end{bmatrix} \begin{pmatrix} x_2 \\ x'_2 \end{pmatrix}, \quad (3)$$

where $L_{23}$ is the beam transmission path length between R4BPM01 and R1BPM01. With Eqs. (2) and (3), the beam position and slope at R1BPM01 $(x_{3t}, x'_{3t})$ in theory can be calculated.

In actual measurement, by using the measurement data of R1BPM01 and R1BPM02 with BC chicane bump, the beam position and slope at R1BPM01 can be calculated. Defining $(x_{3m}, x'_{3m})$ and $(x_4, x'_4)$ as the transverse phase space coordinates of the circular beam at R1BPM01 and R1BPM02, $x_{3m}$ and $x_4$ can be measured in the machine study. The transfer relationship from R1BPM01 to R1BPM02 can be given as

$$\begin{pmatrix} x_4 \\ x_4' \end{pmatrix} = \begin{bmatrix} \sqrt{\frac{\beta_4}{\beta_3}} \times (\cos \Delta\Psi_{43} + \alpha_3 \times \sin \Delta\Psi_{43}) & \sqrt{\beta_4 \times \beta_3} \times \sin \Delta\Psi_{43} \\ \frac{(\alpha_3 - \alpha_4) \times \cos \Delta\Psi_{43} - (1 + \alpha_4 \times \alpha_3) \times \sin \Delta\Psi_{43}}{\sqrt{\beta_4 \times \beta_3}} & \sqrt{\frac{\beta_3}{\beta_4}} \times (\cos \Delta\Psi_{43} - \alpha_4 \times \sin \Delta\Psi_{43}) \end{bmatrix} \begin{pmatrix} x_{3m} \\ x_{3m}' \end{pmatrix}, \quad (4)$$

where $(\alpha_3, \beta_3)$ and $(\alpha_4, \beta_4)$ are the twiss parameters at R1BPM01 and R1BPM02, $\Delta\Psi_{43} = \psi_4 - \psi_3$ while $\psi_3$ and $\psi_4$ are the phases at R1BPM01 and R1BPM02. By using Eq. (4), the beam slope at R1BPM01 can be calculated:

$$x_{3m}' = \frac{x_4 - \sqrt{\frac{\beta_4}{\beta_3}} \times (\cos \Delta\Psi_{43} + \alpha_3 \times \sin \Delta\Psi_{43}) \times x_{3m}}{\sqrt{\beta_4 \times \beta_3} \times \sin \Delta\Psi_{43}}. \quad (5)$$

Therefore, the beam position and slope at R1BPM01 ($x_{3m}$, $x_{3m}'$) in actual measurement can be obtained.

By comparing ($x_{3t}$, $x_{3t}'$) and ($x_{3m}$, $x_{3m}'$), the information about the residual orbit of BC bump can be obtained. If ($x_{3t}$, $x_{3t}'$) and ($x_{3m}$, $x_{3m}'$) are not consistent, the residual orbit of BC bump exists and need to be corrected.

**Correction method to the residual orbit of BC bump**

In the injection system, there are three auxiliary windings of BC magnets (TRIM2 on the magnet BC2, TRIM3 on the magnet BC3, TRIM4 on the magnet BC4) which can be used for correcting the residual orbit of BC bump. By adjusting the three auxiliary windings, while there is a group setting values of magnetic fields which can make ($x_{3t}$, $x_{3t}'$) and ($x_{3m}$, $x_{3m}'$) consistent, the residual orbit of BC bump can be corrected.

Defining $(\Delta B_2', \Delta B_3', \Delta B_4')$ as a group magnetic field values of the three auxiliary windings of BC magnets, then the deflection angles that the circular beam through

$$\Delta\theta_2 = -\frac{\Delta B_2' L_2}{(B\rho)_2}, \quad (6)$$

$$\Delta\theta_3 = -\frac{\Delta B_3' L_3}{(B\rho)_3}, \quad (7)$$

$$\Delta\theta_4 = -\frac{\Delta B_4' L_4}{(B\rho)_4}, \quad (8)$$

where $L_2$, $L_3$, $L_4$ are the effective lengths of magnets BC2, BC3, BC4 and $(B\rho)_2$, $(B\rho)_3$, $(B\rho)_4$ are their magnetic rigidities.

In order to calculate the effects of the three auxiliary windings on the circular beam, defining ($x_{c2}$, $x_{c2}'$), ($x_{c3}$, $x_{c3}'$) and ($x_{c4}$, $x_{c4}'$) as the transverse phase space coordinates of the circular beam at BC2, BC3 and BC4, ($x_{3c}$, $x_{3c}'$) as the transverse phase space coordinates of the circular beam at R1BPM01 with the effects of the three auxiliary windings. The transfer relationship from R4BPM01 to R1BPM01 with the effects of the three auxiliary windings can be given as

$$\begin{pmatrix} x_{c2} \\ x_{c2}' \end{pmatrix} = \begin{bmatrix} 1 & L_{2c2} \\ 0 & 1 \end{bmatrix} \begin{pmatrix} x_2 \\ x_2' \end{pmatrix} + \begin{pmatrix} 0 \\ \Delta\theta_2 \end{pmatrix}, \quad (9)$$

$$\begin{pmatrix} x_{c3} \\ x_{c3}' \end{pmatrix} = \begin{bmatrix} 1 & L_{c23} \\ 0 & 1 \end{bmatrix} \begin{pmatrix} x_{c2} \\ x_{c2}' \end{pmatrix} + \begin{pmatrix} 0 \\ \Delta\theta_3 \end{pmatrix}, \quad (10)$$

$$\begin{pmatrix} x_{c4} \\ x_{c4}' \end{pmatrix} = \begin{bmatrix} 1 & L_{c34} \\ 0 & 1 \end{bmatrix} \begin{pmatrix} x_{c3} \\ x_{c3}' \end{pmatrix} + \begin{pmatrix} 0 \\ \Delta\theta_4 \end{pmatrix}, \quad (11)$$

$$\begin{pmatrix} x_{3c} \\ x_{3c}' \end{pmatrix} = \begin{bmatrix} 1 & L_{4c3} \\ 0 & 1 \end{bmatrix} \begin{pmatrix} x_{c4} \\ x_{c4}' \end{pmatrix}, \quad (12)$$

where $L_{2c2}$ is the beam transmission path length between R4BPM01 and BC2, $L_{c23}$ is the beam transmission path length between BC2 and BC3, $L_{c34}$ is the beam transmission path length between BC3 and BC4, $L_{4c3}$ is the beam transmission path length between BC4 and R1BPM01. In addition, the lengths of the three auxiliary windings of BC magnets are ignored in Eqs. (9), (10), and (11) because they are much smaller than $L_{2c2}$, $L_{c23}$, and $L_{c34}$.

If there is a group magnetic field values of the three auxiliary windings $(\Delta B'_2, \Delta B'_3, \Delta B'_4)$ which can make $(x_{3c}, x'_{3c})$ and $(x_{3m}, x'_{3m})$ consistent, then the three auxiliary windings and the residual orbit of BC bump have the same effects on the circular beam. Therefore, the required magnetic field values of the three auxiliary windings which can correct the residual orbit of BC bump can be obtained:

$$\Delta B_2 = -\Delta B'_2, \ \Delta B_3 = -\Delta B'_3, \ \Delta B_4 = -\Delta B'_4. \qquad (13)$$

With the group setting values of magnetic fields $(\Delta B_2, \Delta B_3, \Delta B_4)$, by adjusting the three auxiliary windings, $(x_{3t}, x'_{3t})$ and $(x_{3m}, x'_{3m})$ can be made consistent. Therefore, the residual orbit of BC bump can be corrected and the circular beam orbit outside the injection region cannot be affected.

**Control program to measure and correct the residual orbit of BC bump**

Based on the XAL application development environment which was developed initially by SNS laboratory [9-11], the control program that measuring and correcting the residual orbit of BC bump had been written, as shown in Fig. 2. It can be found that, while the COD data of the four BPMs is measured, the residual orbit of BC bump can be measured by pressing the "Judgement" button. If the residual orbit of BC bump exists, by pressing the "Calbutton" button, the suitable setting values of magnetic fields $(\Delta B_2, \Delta B_3, \Delta B_4)$ can be calculated. Then, by pressing the "Set Trim Values" button which can adjust the three auxiliary windings of BC magnets, the residual orbit of BC bump can be corrected.

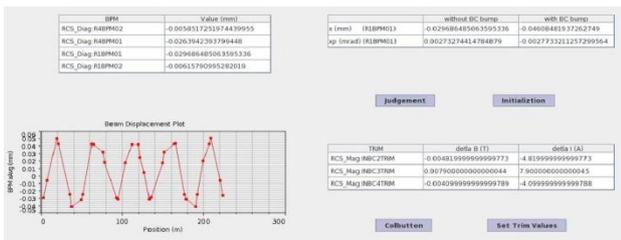

**Fig. 2** Control program that the residual orbit of BC bump can be measured and corrected.

**Numerical simulation**

In the numerical simulation, in order to measure and correct the residual orbit of BC bump, a corrector is necessary in the BC bump region. By adjusting the corrector, the residual orbit of BC bump can be created. Fig. 3 shows the measurement results to the residual orbit of BC bump in the simulation. It can be found that:

$$\begin{cases} x_{3t} - x_{3m} = 1.7 \text{ mm} \\ x'_{3t} - x'_{3m} = 0.5 \text{ mrad} \end{cases},$$

i.e. $(x_{3t}, x'_{3t})$ and $(x_{3m}, x'_{3m})$ are not consistent which means the BC bump is imperfect. Therefore, the residual orbit of BC bump exists and need to be corrected.

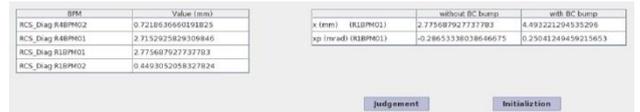

**Fig. 3** Measurement results to the residual orbit of BC bump in the simulation.

By adjusting the three auxiliary windings of BC magnets, the residual orbit of BC bump can be corrected. Fig. 4 shows the correction results to the residual orbit of BC bump in the simulation. It can be found that there is a group magnetic field setting values of the three auxiliary windings which can make

$$\begin{cases} x_{3t} - x_{3m} = 0.0 \text{ mm} \\ x'_{3t} - x'_{3m} = 0.0 \text{ mrad} \end{cases},$$

i.e. ($x_{3t}$, $x'_{3t}$) and ($x_{3m}$, $x'_{3m}$) are consistent. Therefore, the residual orbit of BC bump is corrected and the circular beam orbit outside the injection region cannot be affected.

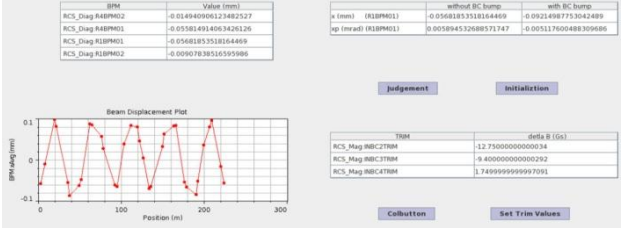

**Fig. 4** Correction results to the residual orbit of BC bump in the simulation.

With the numerical simulation, it can be seen that the methods to measure and correct the residual orbit of BC bump work well and can be applied to the beam commissioning. The measurement accuracy mainly depends on the precisions of the RCS BPMs.

## Machine study

The methods to measure and correct the residual orbit of BC bump were tested and applied to the beam commissioning of CSNS/RCS. In the early stage of the machine study, in order to make the orbit correction of the RCS more precise, the residual orbit of BC bump was measured and corrected. Fig. 5 shows the measurement results to the residual orbit of BC bump in the machine study. It can be found that:

$$\begin{cases} x_{3t} - x_{3m} = 1.1 \text{ mm} \\ x'_{3t} - x'_{3m} = 0.05 \text{ mrad} \end{cases}.$$

The measurement results in the machine study show that the residual orbit of BC bump is not large.

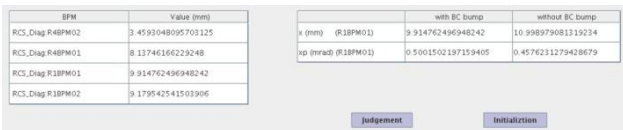

**Fig. 5** Measurement results to the residual orbit of BC bump in the machine study.

By using the three auxiliary windings of BC magnets, the residual orbit of BC bump can be corrected. Fig. 6 shows the correction results to the residual orbit of BC bump. It can be found that, by adjusting the three auxiliary windings of BC magnets, the residual orbit of BC bump can be reduced, i.e.

$$\begin{cases} x_{3t} - x_{3m} = 0.5 \text{ mm} \\ x'_{3t} - x'_{3m} = 0.4 \text{ mrad} \end{cases}.$$

The above correction results show that: $x'_{3t}$ and $x'_{3m}$ are nearly consistent; $x_{3t} - x_{3m}$ is in the range of error due to the precisions of the BPMs （the precisions of the BPMs are around 1 mm [12]）. As a result, the residual orbit of BC bump can be reduced by the above method. Therefore, after the correction of the residual orbit of BC bump, the circular beam orbit outside the injection region was barely affected and the precision of the orbit correction of the RCS was also barely affected.

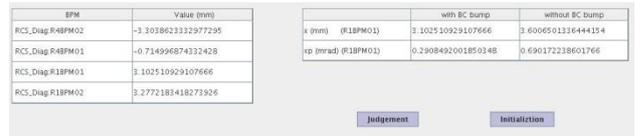

**Fig 6** Correction results to the residual orbit of BC bump in the machine study.

## Summary and discussion

During the beam commissioning of CSNS, the residual orbit of BC bump was an important problem which can affect the orbit correction and beam loss control of the RCS. In this paper, a measurement method to the residual orbit of BC bump was studied and given in detail. Furthermore, by using the three auxiliary windings of BC magnets, a method to correct the residual orbit of BC bump was developed. The numerical simulation results showed that the methods to measure and correct the residual orbit of BC bump had worked well and they can be applied to the beam commissioning. During the machine study, after the data analysis, the research results showed that, within the precisions of the RCS BPMs, the residual orbit of BC bump was not large and can be reduced by the three auxiliary windings of BC magnets.

For the above method to measure and correct the residual orbit of BC bump, there are many errors in systems and measurement, such as the magnetic field errors and power supply errors of related magnets, the precisions of the BPMs, and so on. According to the measurement results of BC magnets [13], the field uniformity of any one BC magnet is smaller than 0.3% and the accuracy of BC power supply is smaller than 0.1%. By calculation, the transfer matrix deviation due to the errors of magnetic fields and power supplies is very small which is ignored in this paper. After preliminary equipment testing and machine study, the precisions of the BPMs are around 1 mm [12]. However, the real error precisions of different BPMs are unknown which need further study in the future.

## Acknowledgments

The authors would like to thank T.G. Xu, J.L. Sun and other CSNS colleagues for the discussions and consultations.